\documentclass{aa}  

\usepackage{natbib}
\usepackage{graphicx}
\usepackage{txfonts}

\usepackage[T1]{fontenc}
\usepackage{ae,aecompl}

\graphicspath{{./Figures/}}
\usepackage{amsmath}    
\usepackage{amssymb}    
\usepackage{mathtools}
\usepackage{array}
\usepackage{color}
\usepackage{bm}
\usepackage{mathrsfs}
\usepackage{dcolumn}
\usepackage{float}

\renewcommand{\epsilon}{\varepsilon}

\def\u{{\bf u}}

\def\nnb{\nonumber\\}

\def\deriv#1#2{\frac{\mathrm{d} #1}{\mathrm{d} #2}}

\def\Gr{\mathcal{G}}
\def \G {{\bf G}}

\def\u{{\bf u}}

\def\u{{\bf u}}                

\def\ub{{\bf u}_\mathrm{b}}
\def\uc{{\bf u}_\mathrm{c}}

\def\uk{{\bf u}_\mathrm{k}}

\begin{document}

        \title{Nearly polar orbit of the sub-Neptune HD3167 c }
        \subtitle{Constraints on the dynamical history of a multi-planet system}
        \titlerunning{Nearly polar orbit of a sub-Neptune}
        \author{S. Dalal
                \inst{1},
                G. H\'ebrard
                \inst{1,2},
                A. Lecavelier des \'Etangs
                \inst{1}
                A. C. Petit
                \inst{3}
                V. Bourrier 
                \inst{4}        
                J. Laskar
                \inst{3}
                P.-C. K\"{o}nig
                \inst{1}
                A.C.M. Correia
                \inst{5,3} 
        }
        \authorrunning{Dalal et al.}
        \institute{Institut d'astrophysique de Paris (IAP), UMR7095 CNRS, Université Pierre \& Marie Curie, 98bis Boulevard Arago, 75014 Paris
                \and
                Observatoire de Haute-Provence (OHP), CNRS, Université d’Aix-Marseille,04870, Saint-Michel-l’Observatoire, France
                \and
                IMCCE, CNRS-UMR 8028, Observatoire de Paris, PSL University, Sorbonne Université, 77 Avenue Denfert-Rochereau, 75014 Paris,
                France
                \and
                Observatoire de l'Université de Geneve, 51 chemin des Maillettes, Versoix, Switzerland.
                \and
                CFisUC, Department of Physics, University of Coimbra, 3004-516 Coimbra, Portugal.
        }
        
        
        
        \abstract
        {}
        {We present the obliquity measurement, that is, the angle between the normal angle of the orbital plane and the stellar spin axis, of the sub-Neptune planet HD3167 c, which transits a bright nearby K0 star.We study the orbital architecture of this multi-planet system to understand its dynamical history. We also place constraints on the obliquity of planet d based on the geometry of the planetary system and the dynamical study of the system.}
        {New observations obtained with HARPS-N at the Telescopio Nazionale Galileo (TNG) were employed for our analysis. The sky-projected obliquity was measured using three different methods:  the Rossiter-McLaughlin anomaly, Doppler tomography, and reloaded Rossiter-McLaughlin techniques. We performed the stability analysis of the system and investigated the dynamical interactions between the planets and the star.  }
        {HD3167 c is found to be nearly polar with sky-projected obliquity, $\lambda$ = -97$^{\circ} \pm$ 23$^{\circ}$. This misalignment of the orbit of planet c with the spin axis of the host star is detected with 97\% confidence. The analysis of the dynamics of this system yields coplanar orbits of planets c and d. It also shows that it is unlikely that the currently observed system can generate this high obliquity for planets c and d by itself. However, the polar orbits of planets c and d could be explained by the presence of an outer companion in the system. Follow-up observations of the system are required to confirm such a long-period companion.}
        {}

        
        \maketitle
        %
        \section{Introduction}
        Obliquity is defined as the angle between the normal angle of a planetary orbit and the rotation axis of the planet host star. It is an important probe for understanding the dynamical history of exoplanetary systems. Solar system planets are nearly aligned and have obliquities lower than 7$^{\circ}$, which might be a consequence of their formation from the protoplanetary disk. However, this is not the case for all exoplanetary systems. Various misaligned systems, that is, $\lambda$ $\succsim$ 30$^{\circ}$ , including some retrograde \citep[$\lambda \sim $180$^{\circ}$, e.g.,][]{Hebrard2008} or nearly polar \citep[$\lambda \sim$ 90$^{\circ}$, e.g.,][]{Triaud2010} orbits have been discovered. These misaligned orbits may result from Kozai migration and/or tidal friction \citep{Nagasawa2008, Fabrycky2007, Guillochon2011, Correia2011}, where the close-in planets migrate as a result of scattering or of early-on interaction between the magnetic star and its disk \citep{Lai2011}, or the migration might be caused later by elliptical tidal instability \citep{Cebron2011}. Another possibility is that the star has been misaligned since the days when the protoplanetary disk was present as a result of inhomogeneous accretion \citep{Bate2010} or a stellar flyby \citep{Batygin2012}.

        Most of the obliquity measurements are available for single hot Jupiters. Some of the smallest planets detected with a Rossiter measurement are GJ 436 b (4.2 $\pm$ 0.2 R$_{\oplus}$) and HAT-P-11 b (4.4 $\pm$ 0.1 R$_{\oplus}$), which are nearly polar \citep{Bourrier2018,Winn2010}, and 55 Cnc e (1.94 $\pm$ 0.04 R$_{\oplus}$), which is also misaligned \citep{Bourrier2014}, although the latest result has been questioned \citep{Lopez2014}. Kepler 408 b is the smallest planet with a misaligned orbit among all planets that are known to have an obliquity measurement \citep{Kamika2019}. A few obliquity detections have been reported for multi-planet systems such as KOI-94 and Kepler 30 \citep{Hirano2012, Albrecht2013, Ojeda2012}, whose planets have coplanar orbits that are aligned with the stellar rotation.
        
        We study the multi-planet system hosted by HD3167. This system includes two transiting planets and one non-transiting planet. \citet{Vanderburg2016} first reported the presence of two small short-period transiting planets from photometry. The third planet HD3167 d was later discovered in the radial velocity (RV) analysis by \citet{Christiansen2017}. \citet{Gandolfi2017} found evidence of two additional signals in the RV measurements of HD3167 with periods of 6.0 and 10.7 days. However, they were unable to confirm the nature of these two signals. Furthermore, \citet{Christiansen2017} did not find any signal at 6 or 10.7 days. The masses of the transiting planets were found to be 5.02 $\pm$ 0.38 $M_{\oplus} $ for HD3167 b, a hot super-Earth, and 9.80$^{+1.30}_{-1.24}$ $M_{\oplus}$ for HD3167 c, a warm sub-Neptune. The non-transiting planet HD3167 d with a mass of at least 6.90 $\pm$ 0.71 $M_{\oplus}$ orbits the star in 8.51 days. The two transiting planets have orbital periods of 0.96 days and 29.84 days and radii of 1.70  R$_{\oplus}$ and 3.01  R$_{\oplus}$, respectively. We measure the sky-projected obliquity for HD3167 c, whose larger radius makes it the most favorable planet for the obliquity measurements. Because the period of planet c is longer than that of planet b,  the data sampling during a given transit is three times better.

        It is difficult to measure the true 3D obliquity, and most methods only access the projection of the obliquity. The sky-projected obliquity for a transiting exoplanet can be measured by monitoring the stellar spectrum during planetary transits. During a transit, the partial occultation of the rotating stellar disk causes asymmetric line profiles that can be detected using different methods such as the Rossiter-McLaughlin (RM) anomaly, Doppler tomography, and the reloaded RM method. These methods use different approaches to retrieve the path of the planet across the stellar disk. This allows us to quantify the  systematic errors related to the data analysis method. The RM anomaly takes into account that asymmetry in line profiles induces an anomaly in the RV of the star \citep{Queloz2000, Hebrard2008}. However, changes in the cross-correlation function (CCF) morphology are not analyzed. Doppler tomography uses the spectral information present in the CCF of the star rather than just their RV centroids. This method entails tracking the full time-series of spectral CCF by modeling the additional absorption line profiles that are superimposed on the stellar spectrum during the planet transit \citep[e.g.,][]{Collier2010, Bourrier2015, Crouzet2017}. This model is then subtracted from the CCFs, and the spectral signature of the light blocked by the planet remains. Finally, the reloaded RM technique directly analyzes the local CCF that is occulted by the planet to measure the sky-projected obliquity \citep[e.g.,][]{Cegla2016, Bourrier2017}. It isolates the CCFs outside and during the transit with no assumptions about the shape of the stellar line profiles.

        The amplitude of the RM anomaly is expected to be below 2 ms$^{-1}$ for HD3167 c. Detecting such a low-amplitude effect is challenging, therefore we decided to determine the robustness and significance of our results using the three different methods described above. The different methods have their respective advantages and limitations. A combined analysis involving the three complementary approaches therefore provides an obliquity measurement that is more robust against systematic effects that are due to the analysis method.
        
        We measure the sky-projected obliquity of HD3167 c using the three methods and finally discuss the dynamics of the system. This paper is structured as follows. We describe the spectroscopic observations during the transit in Section \ref{observations}. The detection of spectroscopic transit followed by the data analysis using the RM anomaly, Doppler tomography, and the reloaded RM is presented in Section \ref{analysis}. We discuss the obliquity of planets b and d from geometry in Section \ref{bdlambda}. We study the dynamics of the system in Section \ref{dynamics} and explore the possible outer companion in Section \ref{outer}. Finally, we conclude in Section \ref{result}. 
        \section{Observations} \label{observations}
        
        We obtained the spectra of HD3167 during the two transits of planet c on 2016 October 1 and 2017 November 23 with the spectrograph HARPS-N with a total of 35 observations and 24 observations, respectively. HARPS-N, which is located at the 3.58 m Telescopio Nazionale Galileo (TNG, La Palma, Spain), is an echelle spectrograph that allows high-precision RV measurements. Observations were taken with resolving power R = 115 000 with 15 minutes of exposure time. We used the spectrograph with one fiber on the star and the second fiber on a thorium-argon lamp so that the observation had high RV precision. The signal-to-noise ratio (S/N) per pixel at 527 nm for the spectra taken during the 2016 transit was 56 to 117 with an average S/N = 87. The 2017 transit was observed in poor weather conditions with S/N values ranging from 34 to 107 with an average S/N = 72. We primarily worked with the 2016 transit data for the reasons explained in Section \ref{2017}.
        
        The Data Reduction Software (DRS version 3.7) pipeline was used to extract the HARPS-N spectra and to cross-correlate them with numerical masks following the method described in \citet{Baranne1996} and \citet{Pepe2002}. The CCFs obtained were fit by Gaussians to derive the RVs and their uncertainties. We tested different numerical masks such as G2, K0, and K5 and also determined the effect of removing some low S/N spectral orders to obtain the CCFs. These tests were performed to improve the data dispersion after the Keplerian fit. The method of fitting a Keplerian is discussed in detail in Section \ref{detection}. Final RVs were obtained from CCFs that we produced using the K5 mask and removing the first 15 blue spectral orders with low S/N. 
        
        The resulting RVs with their uncertainties are listed in Table \ref{table.obs} for the 2016 observations and in Table \ref{table.2017} in Appendix \ref{data} for the 2017 observations. The typical uncertainties were between 0.6 and 1.5 ms$^{-1}$ with a mean value of 0.9 ms$^{-1}$ for the 2016 data. The stellar and planet parameters for HD3167 that we used were taken from Table 1 and from Table 5 of \citet{Christiansen2017}, except for the value of limb-darkening coefficient ($\epsilon$), which was taken from \citet{Gandolfi2017}.
        
        \begin{table}
                
                \caption{Radial velocities of HD3167 measured on 2016 October 1 with HARPS-N}
                \begin{tabular}{l c c}
                        \hline\hline
                        BJD & RV (ms$^{-1}$)& Uncertainty (ms$^{-1}$) \\ 
                        \hline 
                        57663.38879    &    19526.11    &    0.99\\
                        57663.39881    &    19525.6    &    0.89 \\
                        57663.4097    &    19524.96    &    0.96\\
                        57663.42026    &    19525.43    &    0.86 \\
                        57663.43128    &    19525.48    &    0.80\\
                        57663.44191    &    19527.08    &    0.89\\
                        57663.45255    &    19525.54    &    0.80\\
                        57663.463    &    19525.72    &    0.78\\
                        57663.47382    &    19526.8    &    0.69\\
                        57663.48469    &    19525.99    &    0.61\\
                        57663.49535    &    19528.41    &    0.65\\
                        57663.5057    &    19527.32    &    0.66 \\
                        57663.51666    &    19528.51    &    0.69\\
                        57663.52705    &    19529.29    &    0.76\\
                        57663.53812    &    19528.75    &    0.71\\
                        57663.54859    &    19530.09    &    0.66\\
                        57663.5594    &    19529.57    &    0.71\\
                        57663.56994    &    19530.79    &    0.73 \\
                        57663.58084    &    19529.89    &    0.74\\
                        57663.59121    &    19529.66    &    0.75\\
                        57663.60227    &    19531.37    &    0.78\\
                        57663.61288    &    19530.97    &    0.79\\
                        57663.62363    &    19529.69    &    0.83\\
                        57663.63458    &    19530.74    &    0.79\\
                        57663.64483    &    19533.16    &    0.83\\
                        57663.65581    &    19531.99    &    0.85 \\
                        57663.66643    &    19530.85    &    0.77\\
                        57663.67668    &    19532.44    &    0.90\\
                        57663.68756    &    19532.86    &    1.01\\
                        57663.69801    &    19532.29    &    1.30\\
                        57663.70995    &    19530.61    &    1.23\\
                        57663.7196    &    19531.13    &    1.17\\
                        57663.73065    &    19532.3    &    1.16\\
                        57663.74124    &    19532.95    &    1.32 \\
                        57663.75162    &    19532.51    &    1.49 \\

                        \hline 
                        \label{table.obs}
                \end{tabular}
        \end{table}
        
        \section{Analysis} \label{analysis}

        \subsection{Detection of a spectroscopic transit }   \label{detection}
\begin{figure}
                \centering
                \includegraphics[width=9cm,clip=true]{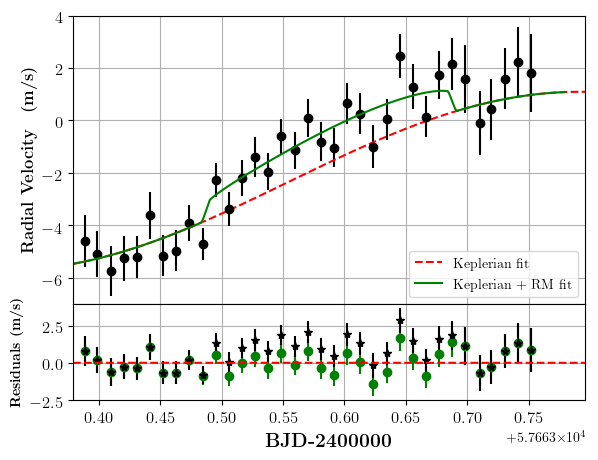}
                \caption{RV measurements of HD3167 taken on 2016 October 1 as function of time. \textit{Upper panel}: Solid black circles represent the HARPS-N data, the dashed red line indicates the Keplerian fit, and the solid green line depicts the final best fit with the RM effect. \textit{Lower panel}: Black solid circles are the residuals after subtracting the Keplerian, and green solid circles are the residuals after subtracting the best-fit RM model.}
                \label{fig.2016}
        \end{figure}

        Figure \ref{fig.2016} displays the RV measurements of HD3167 during the 2016 transit of planet c. The upper panel shows RVs along with the best-fit RM model found from $\chi^{2}$ minimization (discussed in Section \ref{rm}), and the lower panel shows residual RVs after the fit. The red dashed line is the Keplerian model for the orbital motion of the three planets. During the transit, the deviation between the Keplerian model and the observed RVs is caused by the RM anomaly.
        
        To separate the observation taken during the planet transit, only RVs between the beginning of the ingress ($T1$) and end of the egress ($T4$) were considered. The photometric values of mid-transit ($T_{0}$), period ($P$), and transit duration ($T14$) of HD3167 c along with their uncertainties were taken from \citet{Christiansen2017}. The total uncertainty of $\sim$ 16 min on $T_{0}$, inferred from the respective uncertainties of 15 min, 6 min, and 2 min on $P$, $T1$/$T4,$ and $T_{0}$ from \citet{Christiansen2017}, was taken into account in determining the RVs outside the transit. Thirteen RVs (8 before and 5 after the transit) lay outside the transit, while 18 RVs were present inside the transit. Because of the uncertainty in the observed $T_{0}$, it was not clear whether the remaining 4 RVs were present inside or outside the transit. In the following analysis, $T_{0}$ is fixed to the photometric value as the uncertainty on $T_{0}$ is negligible in our analysis, as shown in Section \ref{2016}.
        
        The 13 RVs outside the transit were not sufficient for an independent Keplerian model for the three planets. We therefore took the orbital parameters for the three planets to fit the Keplerian from Table 5 of \citet{Christiansen2017} in Eq. (1) as 
        \begin{equation}\label{key}
        RV= \gamma + \sum_{i=1}^{3} K_{i} \, \big[cos(\textit{f}_{i}+\omega_{i})+e_{i} cos\omega_{i}\big].
        \end{equation}
        
        Here $K_{i}$ represents the RV semi-amplitude, the true anomaly and eccentricity are denoted by $f_i$ and $e_i$ , respectively, and $\omega_i$ is the argument of periastron. Finally, a Keplerian model was fit by minimizing the $\chi^{2}$ considering only one free parameter, that is, the systemic velocity $\gamma$. The average of the residual RVs that were taken outside the transit was found to be 0.11 $\pm$ 0.72 ms$^{-1}$, in agreement with the expected uncertainties.
        
        After the Keplerian fit, we noted that the average of residual RVs inside the transit was 1.17 $\pm$ 0.76 ms$^{-1}$, showing an indication of an RM anomaly detection. We fit this using the RM model in Section \ref{2016}. According to \citet{Gaudi2007}, the expected amplitude of the RM anomaly is 1.7 ms$^{-1}$ , which is within the order of magnitude of the deviation from the Keplerian model observed during the transit.
        
        Furthermore, the slope that is visible in RVs within the observation time (8.7 hours) was due to the short periodicity of HD3167 b ($P_{b}$ = 0.96 day). To compute the mass of HD3167 b, a Keplerian in the RVs outside the transit was fit in which K$_{b}$ was kept as a free parameter. $K_{b}$ was found to be 3.86 $\pm$ 0.35 ms$^{-1}$ , corresponding to a planet mass of HD3167 b of $M_{b}$ = 5.45 $\pm$ 0.50 $M_{\oplus}$. This is consistent with the measurements of \citet{Christiansen2017} ($K_{b}$= 3.58 $\pm$ 0.26 ms$^{-1}$, $M_{b}$ = 5.02 $\pm$ 0.38 $M_{\oplus}$). $K_{b}$ was fixed to the more accurate measurement of \citet{Christiansen2017} in the further analysis.

        We note that the sky-projected obliquity $\lambda$ was defined as the angle counted positive from the stellar spin axis toward the orbital plane normal, both projected in the plane of the sky. The sky-projected obliquity was fit using three different methods, as described in the following sections.

        \subsection{Rossiter-McLaughlin  anomaly} \label{rm}
        The model to fit the RM anomaly is presented in the following section. We applied this model to fit both datasets to measure the sky-projected obliquity.
        
        \subsubsection{Model} 
        
        The method developed by \citet{Ohta2005} was implemented to model the shape of the RM anomaly. These authors derived approximate analytic formulae for the anomaly in RV  curves, considering the effect of stellar limb darkening. Following their approach, we adopted a model with five free parameters: $\gamma$, $\lambda$, the sky-projected stellar rotational velocity $vsini_{\star}$, the orbital inclination $\mathit{i_{p}}$, and the ratio of orbital semi-major axis to stellar radius $a/R_{\star}$. The values of the radius ratio $r_{p} /R_{\star}$, $P$, $T_{0}$ for HD3167 c were fixed to their photometric values \citep{Christiansen2017}, and $\epsilon$ for HD3167 was fixed to 0.54 \citep{Gandolfi2017}. The parameters $\mathit{i_{p}}$ and $a/R_{\star}$ were kept free because their values were poorly constrained from the photometry. Gaussian priors were applied to $\mathit{i_{p}}$ and $a/R_{\star}$ as obtained from photometry \citep{Christiansen2017}. We adopted a value of $vsini_{\star}$ as a Gaussian prior based on the spectroscopy analysis in \citet{Christiansen2017} ($vsini_{\star}$ = 1.7 $\pm$ 1.1 kms$^{-1}$ ). We performed a grid search for the free parameters and computed $\chi^{2}$ at each grid point. The contribution from the uncertainties of $\mathit{i_{p}}$, $a/R_{\star}$ , and $vsini_{\star}$ was also added quadratically to $\chi^{2}$. 
        
        \subsubsection{2016 dataset} \label{2016}
        The data taken on 2016 October 1 are the best dataset for the obliquity measurement in terms of data quality and transit sampling. The 2016 data were fit with the Ohta model, and the reduced $\chi^{2}$ with 30 degrees of freedom (n) for the best-fit model (RM fit) was found to be 0.95. With the RM fit, the averages of residuals inside and outside the transit were 0.01 $\pm$ 0.75 ms$^{-1}$, and 0.11 $\pm$ 0.72 ms$^{-1}$, respectively. The uncertainties agree with the expected uncertainties on the RVs (see Col. 3 of Table \ref{table.obs}). The best-fit value for each parameter corresponds to a minimum of $\chi^{2}$. The 1$\sigma$ error bars were determined for all five free parameters following the $\chi^{2}$ variation as described in \citet{Hebrard2002}. The best-fit values together with 1~$\sigma$ error bars are listed in Table \ref{table.results}. We measured $\lambda$ = $-92^{\circ}\stackrel{+11}{_{-20}}$, indicating a nearly polar orbit.
        
        The derived $vsini_{\star}$ (2.8$\stackrel{+1.9}{_{-1.3}}$ kms$^{-1}$) from the RM anomaly suggested a 2~$\sigma$ detection of the spectral transit. In order to properly determine the significance of our RM detection, we performed Fischer's classical test. The two models considered for the test were a K (only Keplerian) fit and an RM (Keplerian+RM) fit. The $\chi^{2}$ for the K and RM  fits is 63.55 (n = 34) and 28.76 (n = 30), respectively. A significant improvement was noted for the second model with F=1.95 (p=0.03) obtained using an F-test. The improvement to the $\chi^{2}$ was attributed to the RM anomaly detection with 97\% confidence. We conclude that the spectroscopic transit is significantly detected.

        As a test, we applied a similar grid procedure without the spectroscopic constraint on $vsini_{\star}$ from \citet{Christiansen2017}. We obtained $\lambda$ = $-91^{\circ}\stackrel{+7}{_{-16}}$, which is within the 1~$\sigma $ uncertainty. The large $vsini_{\star}$ obtained here (4.8$\pm$ 2.1 kms$^{-1}$) did not significantly affect the measurement of $\lambda$. Because the planetary orbit was found to be polar and it transits near the center of the star (b= 0.50 $\pm$ 0.32, \citep{Christiansen2017}), the corresponding RM anomaly shape did not place a strong constraint on $vsini_{\star}$. The $vsini_{\star}$ can be estimated more accurately using the Doppler tomography technique in Section \ref{dt}. 
        
        Furthermore, the effect of fixed parameters such as $r_{p} /R_{\star}$, $P$, $T_{0}$, $T14,$ and $K_{b}$ on $\lambda$ was investigated. When these fixed parameters were varied within their 1~$\sigma$ uncertainty, $\lambda$ was found to remain within the 1~$\sigma$ uncertainty derived above.

        \begin{figure}
                \centering
                \includegraphics[width=9cm,clip=true]{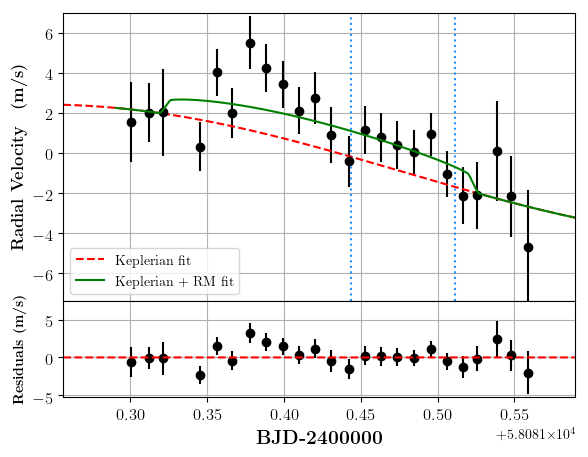}
                \caption{RV measurement of HD3167 taken on 2017 November 23 as a function of time. \textit{Upper panel}: Solid black circles represent the HARPS-N data, the dashed red line indicates the Keplerian fit, and the green line is the over plotted best-fit RM model from the 2016 transit. The blue dotted line marks the transit ingress and egress of planet b. The expected RM amplitude due to the transit of planet b is 0.6 ms$^{-1}$. \textit{Lower panel}: Residuals after the best-fit RM is subtracted.}
                \label{fig.2017}
        \end{figure}
        \subsubsection{2017 dataset} \label{2017}
        Here, we evaluate whether the lower-quality 2017 dataset agrees with the results obtained above using the 2016 dataset. We first determine the observations taken outside the 2017 transit using the same method as explained in Section \ref{detection}. After considering uncertainty on $T_{0}$, we found that only one RV measurement was taken clearly outside the transit. The scarcity of data and poor data sampling outside the transit and along with the low-quality observations during 2017 transit prevented us from finding a good model for a Keplerian and finally an independent value of $\lambda$. Thus the RM model parameters were fixed to the best-fit values from the 2016 transit, and the model derived previously was scaled to the RV level of this epoch. We also realized that during the 2017 transit, HD3167 b and HD3167 c transited simultaneously. However, the expected amplitude of the RM anomaly from HD3167 b is 0.56 ms$^{-1}$ , which is small compared to the RM signal from HD3167 c and the RV measurement accuracy.
        
        Figure \ref{fig.2017} shows the best-fit RM model from Sect. \ref{2016} during the 2017 transit and the residuals after the best-fit RM was subtracted. This fit shows that the 2017 dataset roughly agrees with the results obtained from the RM anomaly fit for the 2016 observations; despite its lower quality, it did not invalidate the results presented in Section \ref{2016}. The residual average inside and outside the transit was found to be 0.23 $\pm$ 1.29 ms$^{-1}$, and 0.39 $\pm$ 1.66 ms$^{-1}$, respectively. The obtained uncertainties were slightly larger than the expected uncertainties on the RVs. The 2017 dataset presented short-term variations in the first half of the transit that could not be due to RM or Keplerian effects. We interpreted them as an artifact due to the bad weather conditions. We achieved no significant improvement from fitting the RM anomaly (F=0.97, p= 0.44), therefore we considered the spectroscopic transit to be not significantly detected in the 2017 data and did not considered it for further analysis.

        \subsection{Doppler tomography} \label{dt}
        Here we present the obliquity measurement we performed on the 2016 dataset using Doppler tomography in order to compare it with the measurement from the RM anomaly technique presented above. When a planet transits its host star, it blocks different regions of the rotating stellar disk, which introduces a Gaussian bump in the spectral lines of the star. This bump can be tracked by inspecting the changes in the CCF, which allows us to measure the obliquity. The stellar rotational speed can also be obtained independent from the spectroscopic estimate by \citet{Christiansen2017}. The CCFs obtained from the DRS with the K5 mask were used for this analysis (Section \ref{observations}). Following the approach of \citet{Collier2010}, we considered a model of the stellar CCF, which is the convolution of limb-darkened rotation profile with a Gaussian corresponding to the intrinsic photospheric line profile and instrumental broadening. When the CCFs are fit by the model including the stellar spectrum and the transit signature, some residual fixed patterns appear that are constant throughout the whole night. These patterns, also called ``sidelobes" by \citet{Collier2010}, are produced by coincidental random alignments between some stellar lines and the lines in the mask when the mask is shifted to calculate the CCFs. To remove these patterns, we assumed that they do not vary during the night, and we averaged the residuals of the out-of-transit CCFs after subtracting the best fit to the CCFs that was calculated by considering the stellar spectrum alone.
        We made a tomographic model that depended on the same parameters as the Keplerian plus RM model (Sect. 3.2), and added the local line profile width, s (non rotating local CCF width) expressed in units of the projected stellar rotational velocity \citep{Collier2010}. The most critical free parameters to fit the Gaussian bump were $\lambda$, $vsini_{\star}$,  $\gamma$, $\mathit{i_{p}}$, $a/R_{\star}$ , and s. Other parameters such as $P$, $r_{p} /R_{\star}$, $T_{0}$ , and $\epsilon$ were fixed to the same values as were used for the RM fit.  
        
        The following merit function was used to fit the CCFs following \citet{Bourrier2015},
        \\
     \begin{equation}
    \chi^{2} = 
    \begin{aligned}
    \sum_{i}^{n_{CCF}}\sum_{j}^{n_{\mu}}\Bigg[\frac{f_{i,j}(model)-f_{i,j}(obs)}{\sigma_{i}}\Bigg]^{2}+\sum_{a_{p}/R_{\star},i_{p}}^{}\Bigg[\frac{x_{tomo}-x_{photo}}{\sigma_{x_{photo}}}\Bigg]^{2} , \\
    \end{aligned}
     \end{equation}

        \begin{figure}
                \centering
                \includegraphics[width=9cm,clip=true]{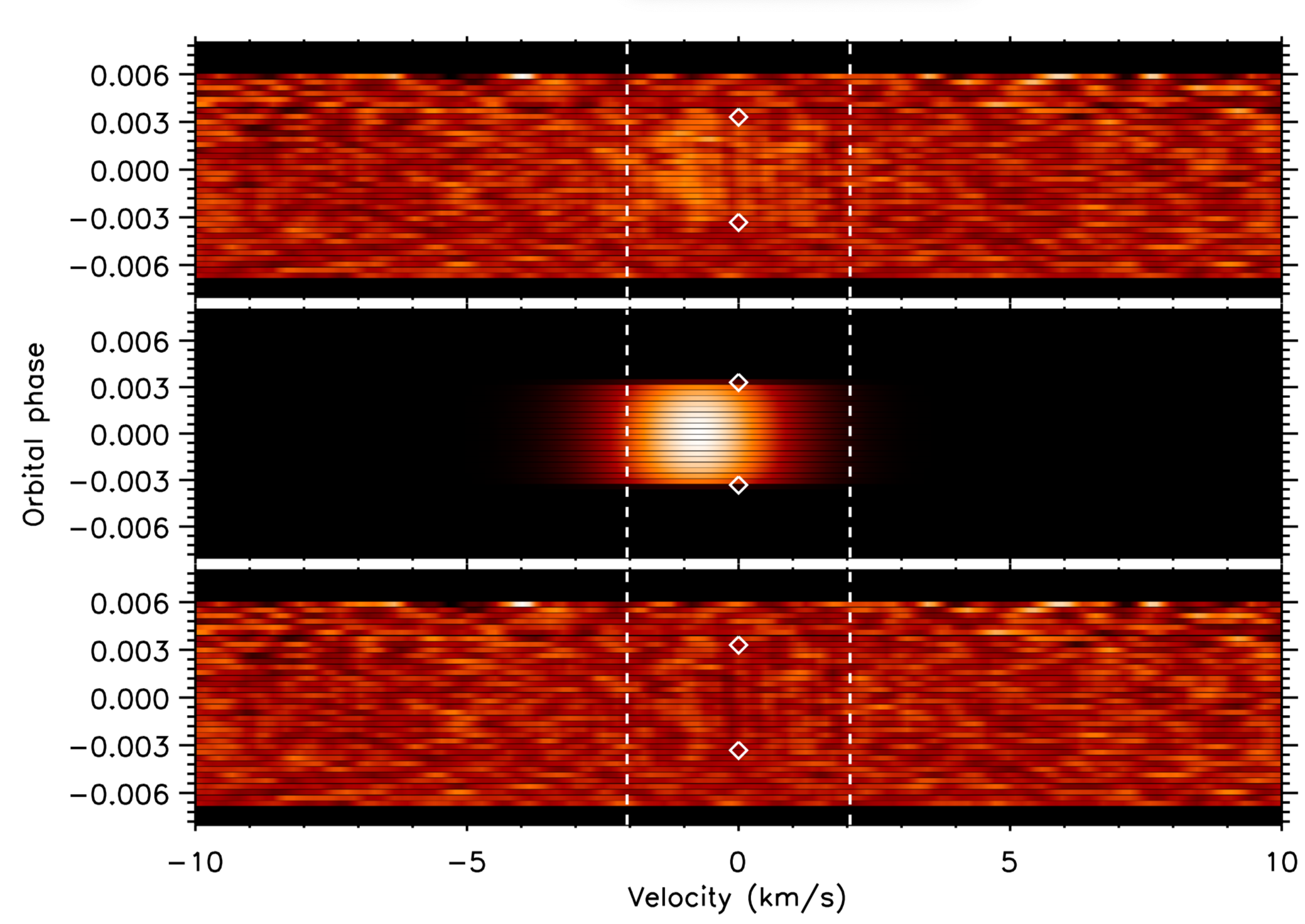}
                \caption{Maps of the time-series CCFs as a function of RV relative to the star (in abscissa) and orbital phase (in ordinate). The dashed vertical white lines are marked at  $\pm vsini_{\star}$, and first and fourth contact of transit is indicated by white diamonds. 
                        \textit{Upper panel}: Map of the transit residuals after the model stellar profile was subtracted. The signature of HD3167 c is the moderately bright feature that is visible from ingress to egress. Middle panel: Transiting signature of HD3167 c using the best-fit model, obtained with $\lambda = - 88^{\circ} $. \textit{Lower panel}: Overall residual map after the model planet signature was subtracted.}
                \label{fig.ddt}
        \end{figure}

        where $f_{i,j}$ is the flux at the velocity point $j$ in the $i_{th}$ observed or model CCFs. The error on the CCF estimate was assumed to be constant over the full velocity range for a given CCF. To find the errors $\sigma_{i}$ in the CCF profiles, we first used the constant errors, which are the dispersion of the residuals between the CCFs and the best-fit model profiles. As the CCFs were obtained using DRS pipeline with a velocity resolution of 0.25 kms$^{-1}$ and the spectra have a resolution of 7.5 kms$^{-1}$, the residuals were found to be strongly correlated. This led to an underestimation of the error bars on the derived parameters. A similar analysis as in  \citet{Bourrier2015} was used to retrieve the uncorrelated Gaussian component of the CCFs. The residual variance as a function of data binning size ($n_{bin}$) is well represented by a quadratic harmonic combination of a white and red noise component,

       \begin{equation}
     \sigma^{2}(n_{bin}) = \Bigg(\bigg(\frac{n_{bin}}{\sigma^{2}_{Uncorr}}\bigg)^{2}+\bigg(\frac{1}{\sigma^{2}_{Corr}}\bigg)^{2}\Bigg)^{-\frac{1}{2}},
       \end{equation}

where $\sigma_{Uncorr}/\sqrt{n_{bin}}$ is the intrinsic uncorrelated noise and $\sigma_{Corr}$ is the constant term characterizing the correlation between the binned pixels. We adopted Gaussian priors for $\mathit{i_{p}}$ and $a/R_{\star}$ from photometry \citep{Christiansen2017}. 
        
        The planet transit was clearly detected in the CCF profiles, as shown in Figure \ref{fig.ddt}. The  $vsini_{\star}$ was found to be 2.1 $\pm$ 0.4 ms$^{-1}$ , which is consistent with the estimate from spectroscopy (vsini$_{\star}$ = 1.7 $\pm$ 1.1 kms$^{-1}$). The sky-projected obliquity was measured to be $\lambda = -88^{\circ} \pm15^{\circ}$ , which is in accordance with the result from the RM analysis (see Section \ref{2016}). Table \ref{table.results} lists the best-fit values together with 1~$\sigma$ error bars.
        
        We also performed a test to check the effect of the fixed parameter $T_{0}$ by varying it within 1~$\sigma$ error bars. The value of $\lambda$ remained within the 1~$\sigma$ uncertainty derived above.

        \begin{figure}
                \centering
                \includegraphics[width=9cm,clip=true]{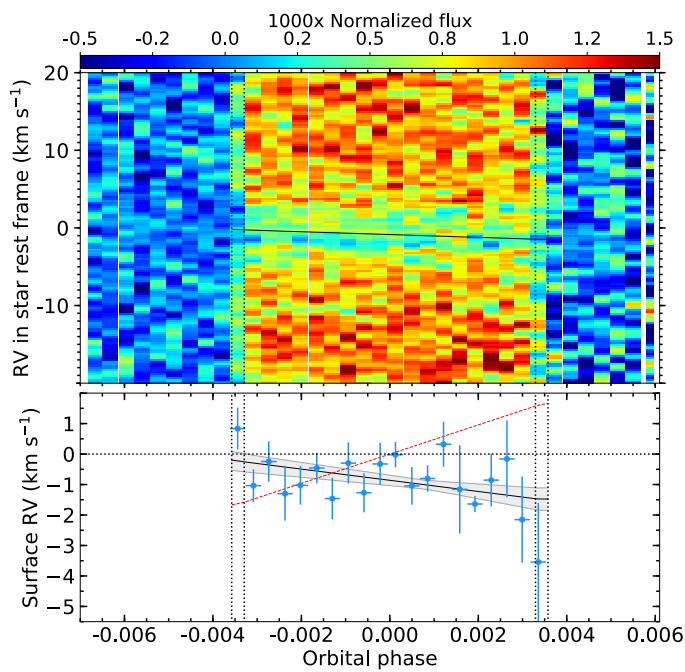}
                \caption{\textit{Upper panel:} Map of the residual CCF series as a function of orbital phase (in abscissa) and RV in the stellar rest frame (in ordinate). Colors indicate flux values. The four vertical dashed black lines show the times of transit contacts. The in-transit residual CCFs correspond to the average stellar line profiles from the regions that are occulted by HD\,3167 c across the stellar disk. The solid black line is the best-fit model to the local RVs of the planet-occulted regions ($\lambda$ = -112.5$^{\circ}$), assuming solid-body rotation for the star ($vsini_{\star}$ = 1.89\,km\,s$^{-1}$). \textit{Lower panel:} RVs of the stellar surface regions occulted by the planet (blue points), best fit with the solid black line (same as in the upper panel). The gray area corresponds to the 1~$\sigma$ envelope of the best fit, derived from the MCMC posterior distributions. The dashed red line shows a model obtained with the same stellar rotational velocity, but an aligned orbit ($\lambda$ = 0$^{\circ}$). This highlights the large orbital misalignment of HD\,3167 c.}
                \label{fig.rrm}
        \end{figure}

        \subsection{Reloaded Rossiter-McLaughlin technique} \label{rrm}
        
        We applied the reloaded RM technique \citep{Cegla2016, Bourrier2018} to the HARPS-N observations of HD3167 c. CCFs computed with the K5 mask (Section \ref{observations}) were first corrected for the Keplerian motion of the star induced by the three planets in the system (calculated with the properties from \citet{Christiansen2017}). The CCFs outside of the transit were co-added to build a master-out CCF, whose continuum was normalized to unity. The centroid of the master-out CCF, derived with a Gaussian fit, was used to align the CCFs in the stellar rest frame. The continuum of all CCFs was then scaled to reflect the planetary disk absorption by HD3167 c, using a light curve computed with the \texttt{batman} package \citep{Kreidberg2015} and the properties from \citet{Christiansen2017}. Residual CCFs were obtained by subtracting the scaled CCF from the master-out (Figure \ref{fig.rrm}). \\
        
        No spurious features are observed in the residual CCFs out of the transit. Within the transit, the residual RM spectrally and spatially resolve the photosphere of the star along the transit chord. The average stellar lines from the planet-occulted regions are clearly detected and were fit with independent Gaussian profiles to derive the local RVs of the stellar surface. We used a Levenberg-Marquardt least-squares minimization, setting flux errors on the residual CCFs to the standard deviation in their continuum flux. Because the CCFs are oversampled in RV, we kept one in four points to perform the fit. All average local stellar lines were well fit with Gaussian profiles, and their contrast was detected at more than 3~$\sigma$ (using the criterion defined by \citet{Allart2017}. The local RV series was fit with the model described in \citet{Cegla2016} and \citet{Bourrier2017}, assuming solid-body rotation for the star. We sampled the posterior distributions of $vsini_{\star}$ and $\lambda$ using the Markov chain Monte Carlo (MCMC) software \texttt{emcee} \citep{Foreman2013}, assuming uniform priors. Best-fit values were set to the medians of the distributions, with 1~$\sigma$ uncertainties derived by taking limits at 34.15\% on either side of the median. The best-fit model shown in Fig 4 corresponds to $vsini_{\star}$ = 1.9 $\pm$ 0.3$\,kms^{-1}$ and $\lambda$ = -112.5$^{\circ}\hspace{-0.1cm}\stackrel{+8.7}{_{-8.5}}$, which agrees at better than 1.4~$\sigma$ with the results obtained from the RM and Doppler tomography (Section \ref{rm} and \ref{dt}). The error bars on $\lambda$ are small because $\mathit{i_{p}}$ and $a/R_{\star}$ were fixed in this particular analysis. However, when $\mathit{i_{p}}$, $T_{0}$, and $a/R_{\star}$ were varied within their 1~$\sigma$ uncertainty, $\lambda$ did not vary significantly and remained within 1~$\sigma$ uncertainty. The best-fit values with their 1~$\sigma$ uncertainties are listed in Table \ref{table.results}.
        
        \begin{table*}[t]

                \caption{Best-fit parameters using three methods} 
                \begin{tabular}{l c c c c}
                        \hline\hline
                        
                        Parameter (unit)& RM fit  &Doppler tomography & reloaded RM & previously published values\\ 
                        \hline 
                        &  & & \\
                       	$\lambda \, (degrees)$  &$ -92.0\stackrel{+11}{_{-20}}$ & $-88 \pm 15$ &$-112.5\stackrel{+8.7}{_{-8.5}}$& - \\ 
                       & & & \\
                       $vsini_{\star} \,(kms^{-1})$& $2.8\stackrel{+1.9}{_{-1.3}} $&$2.1 \pm 0.4$ &$1.9\pm0.3$& $1.7\pm1.1 \, (1)$ \\
                       & & & \\
                       $\gamma\, (kms^{-1})$& $19.5310\stackrel{+0.0003}{_{ -0.0002}}$ & $19.530 \pm 0.009$ & $19.5286 \pm 0.0062 $& - \\
                       & & & \\
                       $\mathit{i_{p}} \,(degrees)$ & $89.5\stackrel{+0.3}{_{-1.2}}$ & $88.91 \pm 0.6$ & $89.3^{\star} $ &$89.3\stackrel{+0.5}{_{-0.96}} (1)$ \\
                       & & & \\
                       $a/R_{\star}$ & $43.3\stackrel{+3.5}{_{-16.0}}$ & $36 \stackrel{+10}{_{-7}}$& $40.323^{\star} $&$40.323\stackrel{+5.55}{_{-12.62}} (1)$  \\
                        & & & \\
                
                        \hline
                        \label{table.results}
                        
                \end{tabular}\\
                $\star$:  Fixed to their photometric value~\citep{Christiansen2017}\\
                \tablebib{
                        (1)~\citet{Christiansen2017}
                }

        \end{table*}

        \subsection{Comparison between the three methods}
        The most commonly used method to estimate sky-projected obliquity using RV measurements is the analysis of the RM anomaly. However, the RM method does not exploit the full spectral CCF. In some extreme cases, the classical RM method can introduce large biases in the sky-projected obliquity because of asymmetries in the local stellar line profile or variations in its shape across the transit chord \citep{Cegla2016a}. The Doppler tomography method is less affected than the RM anomaly method because it explores the full information in the CCF. However, a bias in the obliquity measurements can also be introduced by assuming a constant, symmetric line profile and ignoring the effects of the differential rotation. Results from the reloaded RM technique suggest that the bias is not significant here. The reloaded RM technique does not make prior assumptions of the local stellar line profiles and allows a clean and direct extraction of the stellar surface RVs along the transit chord. This results in an improved precision on the obliquity, albeit under the assumption that the transit light-curve parameters (in particular the impact parameter and the ratio of the planet-to-star radius) are known to a good enough precision to be fixed. In the present case, we might thus be underestimating the uncertainties on $\lambda$ with this method.
        
        The sky-projected obliquities measured by all three methods agree to better than 1.4~$\sigma$. This confirms that the spectroscopic transit in the 2016 data is significantly detected and suggests that the corresponding obliquity measurement is not reached by strong systematics that would be due to the method. Combining the $\lambda$ values from all three methods, we estimated the sky-projected obliquity for HD3167 c to be $\lambda$ = -97$^{\circ}$ $\pm$ 23$^{\circ}$, after taking into account both the systematic and statistical errors. We adopted this conservative value in our final obliquity measurement.
        
        As discussed in Section \ref{2016}, the stellar rotation speed was poorly constrained by the RM method. However, the $vsini_\star$ more accurately measured from Doppler tomography and the reloaded RM technique was consistent with the measurements of \citet{Christiansen2017}. The $vsini_\star$ from three methods was also found to be within 1~$\sigma$. Furthermore, the two photometric parameters $a/R_\star$ and i$_{p}$  also agreed  within their uncertainties for the RM and Doppler tomography methods. The systemic velocity $\gamma$ is slightly different in each case because a different definition was employed in each method.

        \section{Obliquity of planets b and d from geometry} \label{bdlambda}
        The spectroscopic transit observations gave constraints only on the obliquity of planet c. Although planet b is also transiting, the low amplitude for the RM signal during the transit precludes measuring its obliquity with the present data. However, because both planets b and c are transiting planets, the mutual inclination can be constrained. 
        \begin{figure}
                \centering
                \includegraphics[width=9cm]
                {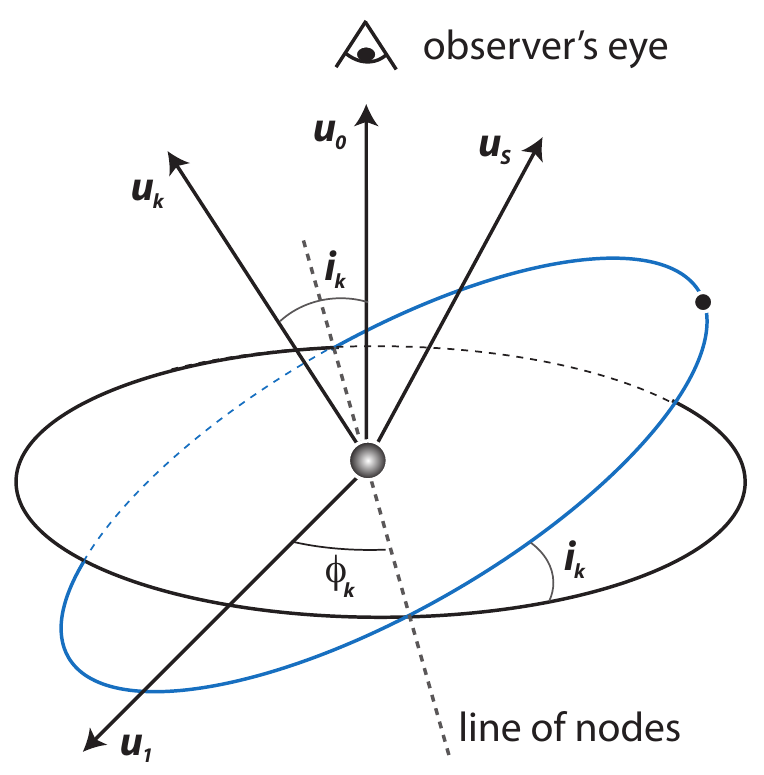}
                \caption[]{
                        Pictorial representation of the reference angles and the unit vectors. $\u_\mathrm{S}$ corresponds to the direction of the stellar spin.
                }
                \label{vector}
        \end{figure}
        
\begin{figure}
                \centering
                \includegraphics[width=9cm]
                {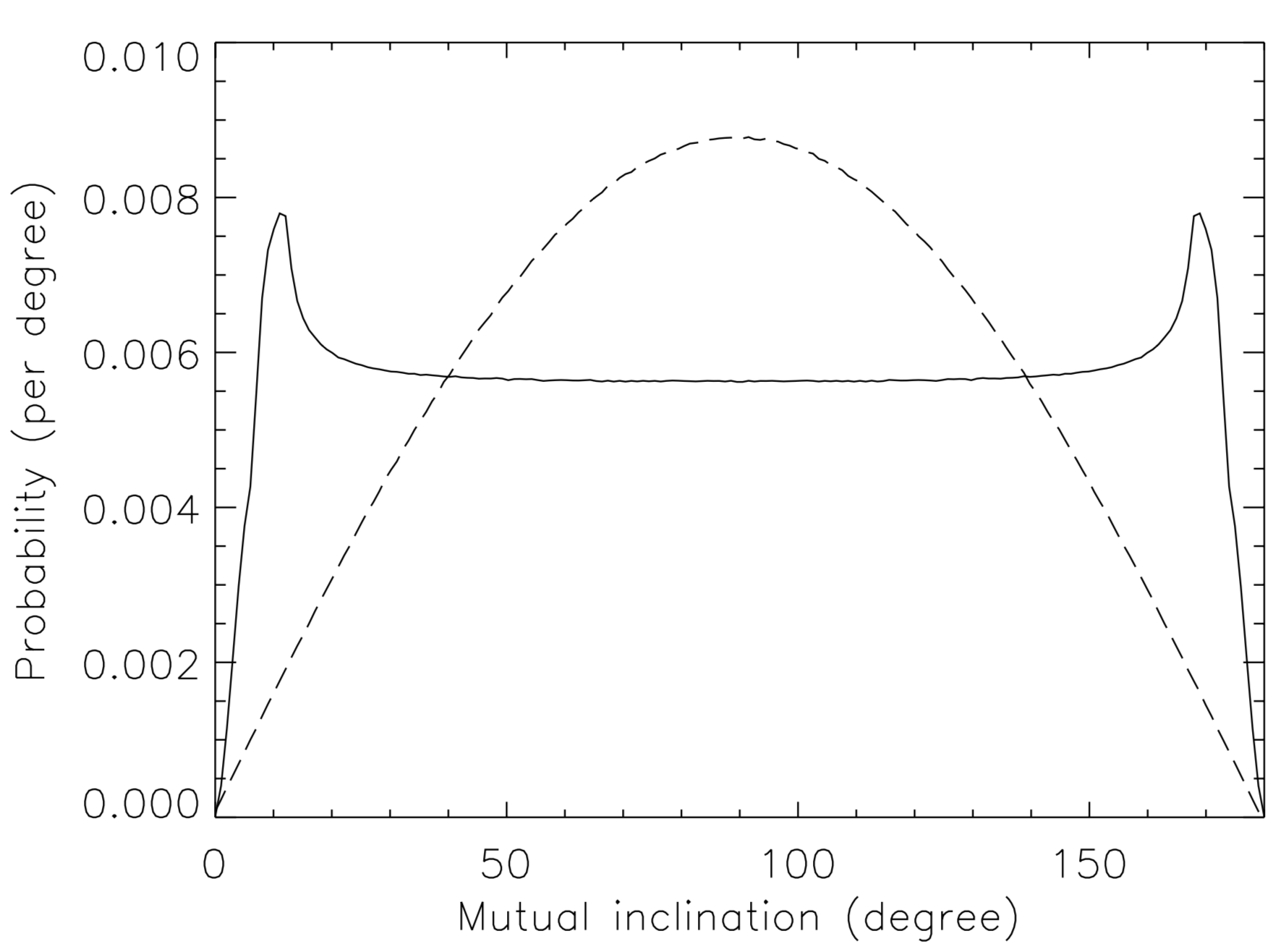}
                \caption[]{
                        Probability distribution of the mutual inclination between the planets b and c (solid line). For comparison, the dotted line shows the probability distribution when neither planet transits.
                }
                \label{fig.mutualbc}
        \end{figure}

        We denote by $\u_0$ the unit vector along the line of sight directed toward Earth and $\u_1$ a unit vector perpendicular to $\u_0$, that is, in the plane of the sky (see Figure \ref{vector}). The orbital planes of planets b and c are characterized by the perpendicular unit vectors $\ub$ and $\uc$. The inclination of their orbits, $i_b$ and $i_c$, is constrained to be $i_b=83.4^{\circ}\hspace{-0.1cm}\stackrel{+4.6}{_{-7.7}}$ and $i_c=89.3^{\circ}\hspace{-0.1cm}\stackrel{+0.5}{_{-0.96}}$ \citep{Christiansen2017}. For a planet $k$ (here $k$ stands for either b or d), we define $\phi_k$ as the angle between $\u_1$ and the projection of $\uk$ on the plane of the sky (this is equivalent to the longitude of the ascending node in the plane of the sky). With these definitions, the mutual inclination between the planets b and c, $i_{bc}$, is given by  
        \begin{equation}
        \cos i_{bc}= \cos i_b \cos i_c + \sin i_b \sin i_c  \cos (\phi_b-\phi_c)
        .\end{equation}
        With cos $i_b$ and cos $i_c$ uniformly distributed within their 1~$\sigma$ error bars and assuming that $\phi_b$ and $\phi_c$ are uniformly distributed between 0 and 2$\pi$, we calculated the probability distribution of $i_{bc}$ (Fig.~\ref{fig.mutualbc}). The probability distribution was found to be close to a uniform distribution, except that it is low for $i_{bc}$ below 10$^{\circ}$ and above 170$^{\circ}$. Based on geometry, no information on the obliquity of planet b can therefore be derived from our measurement of the obliquity of planet c. We note that in the case of two non-transiting planets, the probability distribution of $i_{bc}$ would peak around 90$^\circ$ , as shown by the dotted line in Figure \ref{fig.mutualbc}. 
        
        Planet d would transit  if the condition
        \begin{equation}
        i_{dc} \leq \arctan\left(\frac{R_{\star}}{a_d}\right) - (90-i_c)
        \label{transi.cond}
        \end{equation}
        were fulfilled, where $R_{\star}$ is the stellar radius and $a_d$ is the semi-major axis of planet d. Because planet d does not transit, the mutual inclination between planets c and d must be at least 2.3$^\circ$.
        
        As a result, the obliquity of planets b and d cannot be constrained well from the geometry of the planetary system alone. We place constraints on the obliquity of planet d from the dynamics of the planetary system in Section \ref{dynamics}.

        \section{Dynamics} \label{dynamics}
        
 We study the dynamics of the system to investigate the interactions between planets and stellar spin which could explain the polar orbit of planet c. We also perform the Hill stability analysis to set bounds on the obliquity of planet d in the following section.
        \subsection{Planet mutual inclinations} \label{planetmutual}
        
        \begin{figure}[!b]
                \centering
                \includegraphics[width=9cm]{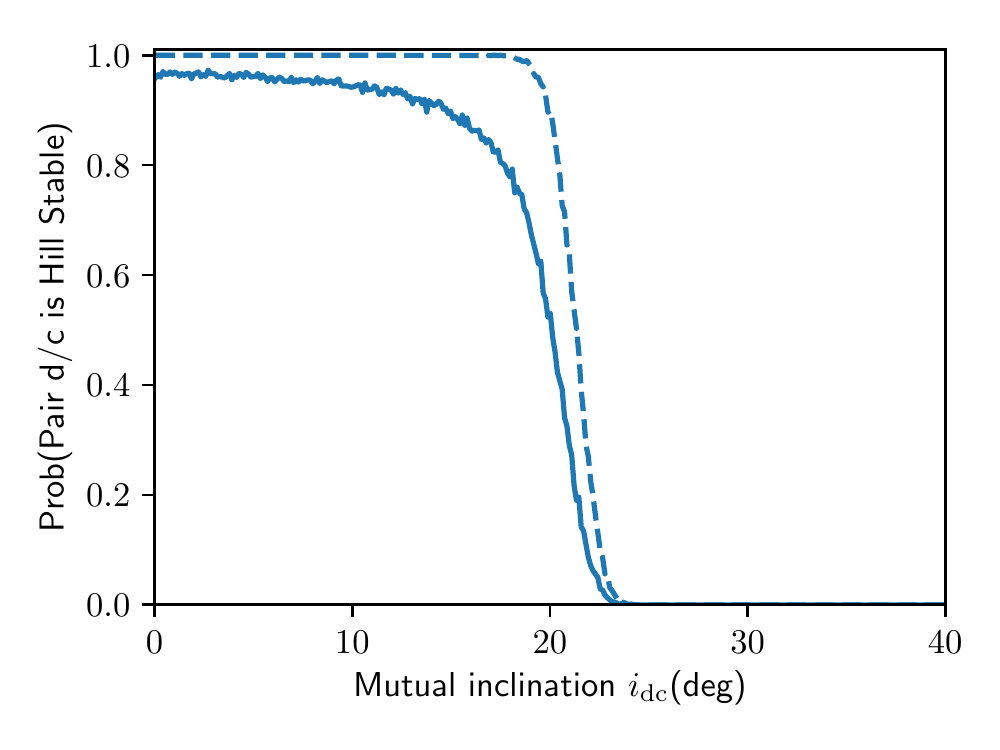}
                \caption{Probability of the pair d-c to be Hill-stable as a function of the mutual inclination of d and c, assuming planet b is within the invariant plane. The masses, semi-major axes, and eccentricity are drawn from the best-fit distribution \citep{Christiansen2017}. The dashed curve corresponds to a system where every planet is on a circular orbit.}
                \label{fig.hill_stable}
        \end{figure}
        While the available observations were unable to geometrically constrain the mutual inclination of the planets, a bound is given by the stability analysis of the system.
Short-period planets with an aligned orbit such as KELT-24 b and WASP-152 b \citep{Rod2019,Santerne2016}, or with an misaligned orbit such as Kepler-408 b and GJ436 b \citep{Kamika2019,Bourrier2018} have been detected. The obliquity distribution of short-period planets is not clear. However, because planet b is close to the star, its orbit is most likely circular and its inclination is governed by the interaction with the star, as shown in Appendix \ref{app.planetb}. The exact inclination of planet b is not important from a dynamical point of view, and it is safe to neglect the influence of planet b when the stability of the system is investigated.
        We focus here on the outer pair of planets to constrain the system and study the simplified system that is only composed of the star and the two outer planets. Our goal is to determine the maximum mutual inclination between planets d and c such that the outer pair remains Hill-stable \citep{Petit2018, Marchal1982}. 
        We first created $10^6$ realizations of the HD3167 system by drawing from the best fit of masses, eccentricities, and semi-major axis 
        distributions given by \citet{Christiansen2017}. To each of these copies of the system, we set the mutual inclination between planets c and d with uniformly spaced values of between $0^{\circ}$ and $ 90^{\circ}$.

       We assumed that the orbit of planet b is in the invariant plane, that is, the plane perpendicular to the angular momentum vector of the whole system
      \footnote{We assumed that planet b is within the invariant plane in order to be able to compute the AMD as a function of the mutual inclination $i_\mathrm{dc}$. Nevertheless, the actual planet b's inclination has little influence on the stability of the pair d-c.}.  As a result, we computed the inclinations $i_\mathrm{c}$ and $i_\mathrm{d}$ with respect to the invariant plane because the projection of the angular momentum onto the invariant plane gives
        \begin{equation}
        G_\mathrm{c}\sin(i_\mathrm{c}) = G_\mathrm{d}\sin(i_\mathrm{d}),
        \end{equation}
        where $G_k=m_k\sqrt{\mathcal{G}M_Sa_k(1-e_k^2)}$ is the norm of the angular momentum of planet $k$. Then, we computed the total angular momentum deficit \citep[AMD,][]{Laskar1997} of the system
        \begin{equation}
        C = \sum_{k=\mathrm{b,c,d}} m_k\sqrt{\Gr M_s a_k}\left(1-\sqrt{1-e_k^2}\cos(i_k)\right),
        \end{equation}
        and we determined whether the pair d-c is Hill-stable. To do so, we compared the AMD to the Hill-critical AMD of the pair \citep[Eq. 30,][]{Petit2018}. We plot in Figure \ref{fig.hill_stable} the proportion of the Hill-stable system binned by mutual inclination $i_\mathrm{dc}$. We also plot the proportion of the Hill-stable pair for a system with circular orbit.

        We observe that for an inclination $i_\mathrm{dc}$ below $21^\circ$, the system is almost certainly Hill-stable. This means that for any orbital configuration and masses that are compatible with the observational constraints, the system will be long-lived with this low mutual inclination.
        We emphasize that long-lived configurations with higher mutual inclination than $21^{\circ}$ exist. \citet{Christiansen2017} gave the example of Kozai-Lidov oscillations with initial mutual inclinations of up to $65^\circ$. However, the choice of initial conditions is fine-tuned because of the circular orbits (a configuration that is rather unlikely for such dynamically excited systems).
        
        When we assume that the stellar spin is aligned with the total angular momentum of the planets, the planet obliquity corresponds to the planet inclination with respect to the invariant plane. When we assume $i_\mathrm{dc}<21^\circ$, the maximum obliquity of planet c is about $9^\circ$. Even if the mutual inclination $i_\mathrm{dc}=65^\circ$, the obliquity only reaches $32^\circ$. Thus, the observed polar orbit shows that the stellar spin cannot be aligned with the angular momentum of the planet. 
        
        From Section \ref{bdlambda} and the previous paragraphs, we deduce that the most likely value for $i_\mathrm{dc}$ is between $2.3^\circ$ and $21^\circ$. Because the mutual inclination of planets c and d is low, we can conclude that planet d is also nearly polar.
        
        \subsection{Interactions of planets and stellar spin}
        
        Because the system's eccentricities and mutual inclinations are most likely low to moderate, we considered the interaction between the stellar spin and the planetary system.
        In particular, we investigated whether the motion of the planets can effectively tilt the star up to an inclination that could explain the polar orbit of planets c and d.
        The currently known estimate of \citet{Christiansen2017} of the stellar rotation period is $27.2 \pm 7$ days, but the period may have slowed down by a factor 10 \citep{Bouvier2013}.
        In order to investigate the evolution of the obliquity that could have occurred in earlier stages in the life of the system, we studied the planet-star interaction as a function of the stellar rotation period.
        
        To do so, we applied the framework of the integrable three-vector problem to the star and the angular momenta of planets d and c \citep{Boue2006, Boue2014, Correia2015}. This model gives both the qualitative and quantitative behavior of the evolution of three vectors that represent different angular momenta directions $\u_\mathrm{S}$, $\u_\mathrm{d}$  , and $\u_\mathrm{c}$ under their mutual interactions. We describe the model in Appendix \ref{app.3vmod}.
        
        As shown in \citet{Boue2014}, the mutual interactions of the three vectors can be described by comparing the different characteristic frequencies\footnote{
       	The characteristic frequencies designate the coupling parameters between the different vectors, as explained in Appendix \ref{app.3vmod}. They have the dimension of a frequency, but are not properly speaking the frequencies of the system. 
       	Here, we use the terminology introduced in \citet{Boue2014}.} 
       of the system $\nu^\mathrm{d/S},\nu^\mathrm{S/d},\nu^\mathrm{d/c}$ , and $\nu^\mathrm{c/d}$ with the expressions given in Eq. \eqref{eq.freqd}.
       The frequency $\nu^{j/k}$ represents the relative influence of the body $j$ over the motion of $\u_k$. In other words, if $\nu^{k/j}\ll\nu^{j/k}$, $\u_j$ is almost constant while $\u_k$ precesses around.
       We here neglect the interactions between the star and planet c versus the interaction between the star and planet d because they are smaller by two orders of magnitude. 
        \begin{figure}
                \centering
                \includegraphics[width=9cm]{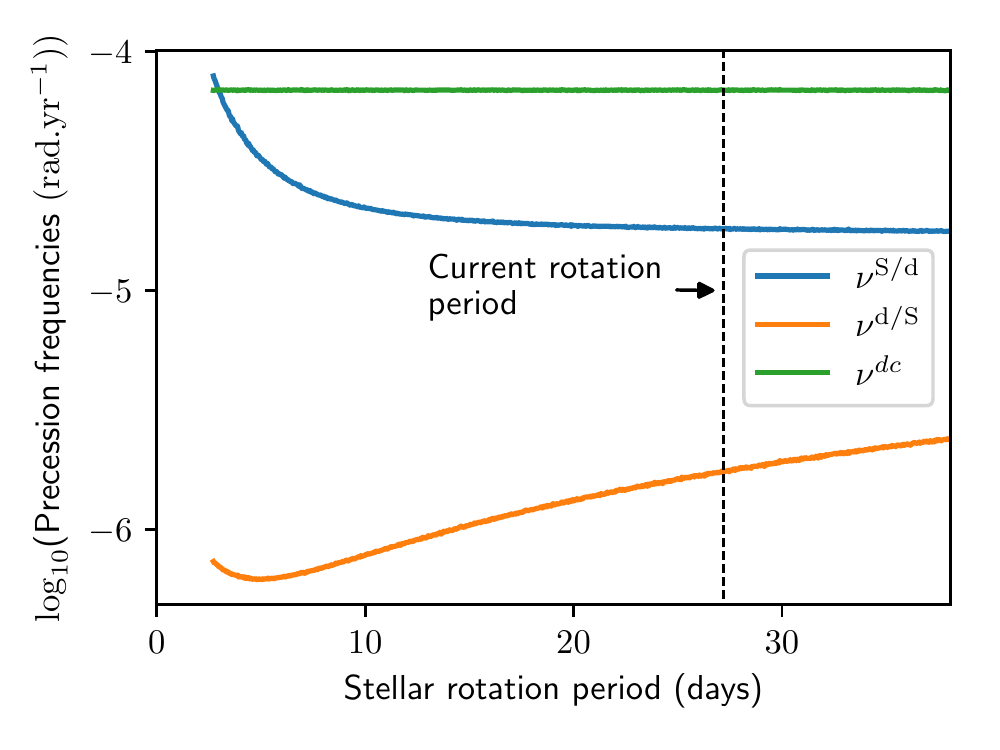}
                \caption{Characteristic frequencies defined in Eq. \eqref{eq.freqd} as a function of stellar period. The current estimated stellar rotation period is marked with a vertical dashed line. The two terms $\nu^\mathrm{d/c}$ and $\nu^\mathrm{c/d}$ are merged into a single curve $\nu^\mathrm{dc}$ because they are almost equal.}
                \label{fig.precd}
        \end{figure}

        Because it is coupled with the star, planet b acts as a bulge on the star that enhances the coupling between the orbits of the outer planets and the star (see Appendix \ref{app.planetb}). We limit our study to the configuration where the strongest coupling occurs, that is, when the orbit of planet b lies within the stellar equatorial plane. The influence of planet b modifies the characteristic frequencies $\nu^\mathrm{d/S}$ and $\nu^\mathrm{S/d}$ , as we show in Appendix \ref{app.planetb}.
        The model is valid in the secular approximation if the eccentricities of planets d and c remain low such that  $G_\mathrm{d}$ and $G_\mathrm{c}$ are constant. \citet{Boue2006} showed that the motion is quasi-periodic. It is possible to give the maximum spin-orbit angle of planet c as a function of the initial inclination of planet d.

        Using the classification of \citet{Boue2014}, we can determine the maximum misalignment between $\u_\mathrm{S}$ and $\u_\mathrm{c}$ as a function of the initial inclination between  $\u_\mathrm{S}$ and $\u_\mathrm{d}$.
        We plot the frequencies (cf. Eq. \eqref{eq.freqd}) as a function of the stellar rotation period in Figure \ref{fig.precd}. We merged the curves that represent $\nu^\mathrm{d/c}$ and $\nu^\mathrm{c/d}$ into $\nu^\mathrm{dc}$ because the two terms are almost equal.
        
        \label{sec.tilt}

        We are in a regime where ${(\nu^\mathrm{d/c}\sim\nu^\mathrm{c/d})\gg (\nu^\mathrm{d/S},\nu^\mathrm{S/d})}$ and the orbital frequencies dominate the interactions with the star. For the shorter periods we have ${(\nu^\mathrm{d/c}\sim\nu^\mathrm{c/d}\sim \nu^\mathrm{S/d})\gg \nu^\mathrm{d/S}}$.
        Nonetheless, in both cases the dynamics are purely orbital, however, meaning that\emph{} the star acts as a point mass and is never coupled with the orbits of the outer planets. It is not possible for planets c and d to reach a high mutual inclination with the stellar spin axis starting from almost coplanar orbits or an even moderate inclination. When planet b is misaligned with the star, it is even harder for the planets to tilt the star.
        
        We conclude that even if the star has had a shorter period in the past, it is unlikely that the currently observed system can by itself generate such a high obliquity for planets~c and d.
        However, high initial obliquities are almost conserved, which means that the observed polar orbits are possible under the assumptions made, even though they are not explained by this scenario.

        \subsection{System tilt due to an unseen companion} \label{dynamicsouter}
        
        We now assume that while the system only presents moderate inclinations, a distant companion on an inclined orbit exists. We consider the configurations that can cause the system to be tilted with respect to the star.
        Once again, we used the framework of \citet{Boue2014}. We considered the vectors $\u_\mathrm{S},\ \u,$ and $\u'$ that give the direction of the stellar spin axis $\mathbf{S}$, the total angular momentum of the planetary system $\G,$ and the angular momentum of the companion $\G'$, respectively.
        The outer companion is described by its mass $m'$, its semi-major axis $a'$, its semi-minor axis $b' = a'\sqrt{1-e'^2}$ , and its initial inclination $I_0$ with respect to the rest of the system, which is assumed to be nearly coplanar or to have moderate inclinations.
        Moreover, we assumed that $\G$ is initially aligned with $\mathbf{S,}$ while the companion is highly inclined with respect to the planetary system, that is to say,  $I_0$ is larger than $45^\circ$ up to $90^\circ$. 
        According to \citet{Boue2014}, all interactions between planets cancel out because we only consider the dynamics of their total angular momentum $\G$.
        
        As in the previous part, we can compare the different characteristics frequencies of the system $\nu^\mathrm{pla/S},\nu^\mathrm{S/pla}, \nu^\mathrm{comp/pla}$ , and $\nu^\mathrm{pla/comp}$ of expression given in Eqs. \eqref{eq.freqcomp} and \eqref{eq.couplingcomp}.
        The companion effectively tilts the planetary system as a single body if its influence on planet c is weaker than the interaction between planets d and c.
        In the other case, planet c will enter Lidov-Kozai oscillations, which can lead to the destabilization of the system through the interactions with planet d.
        \cite{Boue2014} reported the limit at which the outer companion starts to perturb the planetary system and excites the outer planet through Kozai-Lidov cycles.
        They explained that if the coefficient $\beta_{\mathrm{KL}}$ is defined as
        \begin{equation}
        \label{eq.KL}
        \beta_{\mathrm{KL}} = \frac{m'}{m_\mathrm{d}}\left(\frac{a_\mathrm{c}}{a_\mathrm{d}}\right)^2 \left(\frac{a_\mathrm{c}}{b'}\right)^3\end{equation}
        and verifies $\beta_{\mathrm{KL}}\ll 1$, the companion's influence does not perturb the system and tilts it as a whole.
        \begin{figure}
                \centering
                \includegraphics[width=9cm]{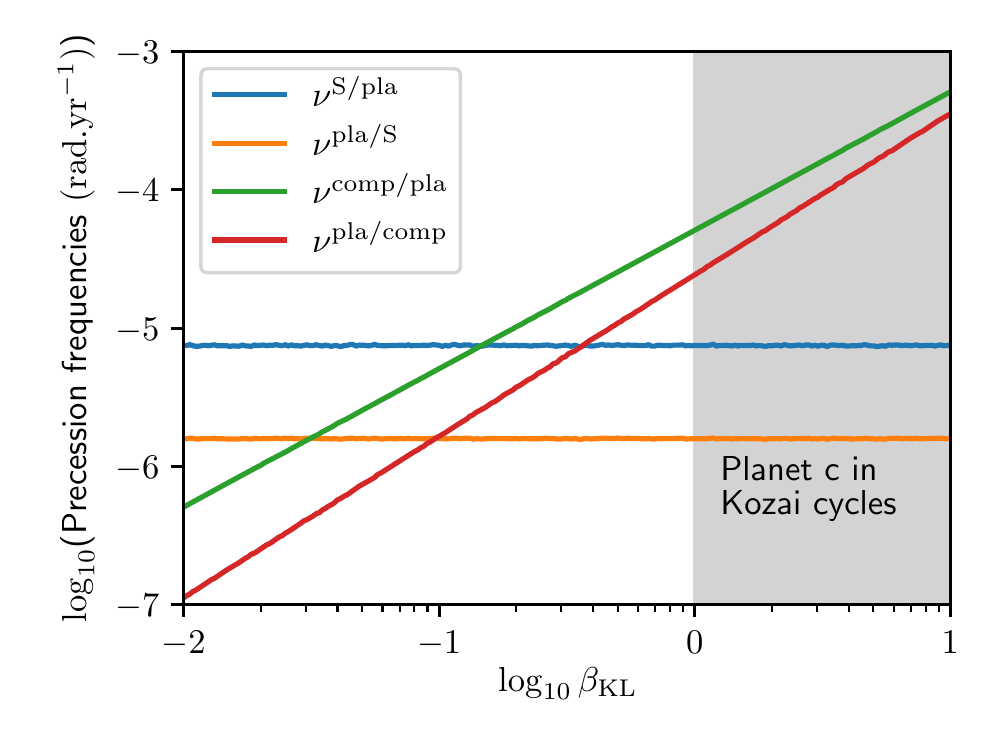}
                \caption{Characteristic frequencies defined in Eq. \eqref{eq.freqcomp} as a function of $\beta_{\mathrm{KL}}$ (see Eq. \ref{eq.KL}). For $\beta_{\mathrm{KL}}>1$, the outer companion can destabilize the observed system.}
                \label{fig.preccomp}
        \end{figure}
        
        We plot in Figure \ref{fig.preccomp} the frequencies $\nu^\mathrm{pla/S},\ \nu^\mathrm{S/pla},\ \nu^\mathrm{comp/pla}$ , and $\nu^\mathrm{pla/comp}$ as a function of $\beta_{\mathrm{KL}}$ and observe different regimes.
        In the first regime, we have $\beta_{\mathrm{KL}}<0.1$ and  $\nu^\mathrm{pla/comp}\ll \nu^\mathrm{comp/pla}\ll \nu^\mathrm{S/pla}$. The influence of the companion is too weak to change the obliquity of the planetary system. For $\beta_{\mathrm{KL}}>1$,  we have $(\nu^\mathrm{pla/comp}\sim \nu^\mathrm{comp/pla})\gg \nu^\mathrm{S/pla}$ , in which regime the system obliquity can reach $I_0$. However, the companion destabilizes the orbit of planet c, which can lead to an increase in eccentricity and mutual inclination between the planets.
        For $0.1\lesssim\beta_{\mathrm{KL}}\lesssim1$, we remark that  $\nu^\mathrm{pla/S}\ll(\nu^\mathrm{pla/comp}\sim \nu^\mathrm{comp/pla}\sim \nu^\mathrm{S/pla})$.
        According to \citeauthor{Boue2014}'s classification, the maximum possible inclination between the star and the planet, that is, their obliquity, is almost twice $I_0$ for $I_0\lesssim 80^\circ$.
        In this regime, an unseen companion can explain the observed polar orbits of HD3167 c and d.
        
        We conclude that some stable configurations with an additional outer companion may explain the high obliquity of planets c and d. We further discuss the possible presence of outer companion signals in the existing RV data in Section \ref{outer}. Accurate measurement of the eccentricities of planets d and c will also help to constrain this scenario better. 
        
        \section{Outer companion} \label{outer}
        To find the possible signatures of an outer companion, we performed two different tests on the RV data from \citet{Christiansen2017}, which cover a span of five months. First we obtained the residual RV after we removed the Keplerian signal caused by the three planets. In the analysis performed by \citet{Christiansen2017}, the linear drift was fixed to 0 ms$^{-1}$yr$^{-1}$ before the Keplerian was fit. However, we detected a linear drift of about 7.6 $\pm$ 1.6 ms$^{-1}$yr$^{-1}$ in the residual velocities. When we assume a circular orbit for the outer companion, this linear drift corresponds to a period of at least 350 days and a mass of at least  0.1 M$_{Jup}$. A body like this has a $\beta_{\mathrm{KL}} \simeq 0.08$, which makes it unlikely that it is able to incline planets c and d with respect to the star.
        
\begin{figure} 
                \centering
                \includegraphics[width=9cm]{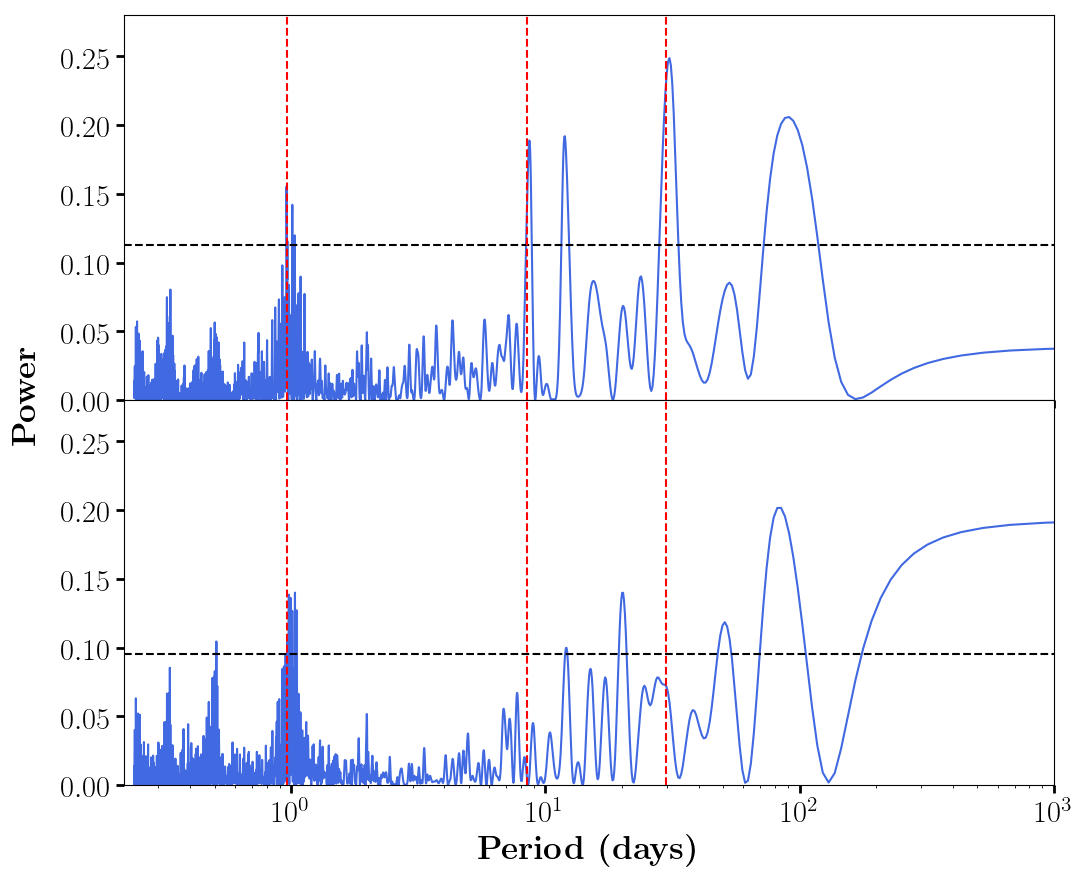}
                \caption{\textit{Upper Panel}: Lomb-Scragle periodogram of the RV of HD3167 taken from \citet{Christiansen2017}. The black dashed line represents the false-alarm probability at 0.1\%, and the three vertical red dashed lines correspond to the periods of the three planets that are currently detected around HD3167. \textit{Lower Panel}: Lomb-Scragle periodogram of the RV data after the three known periodic signals are removed.}
                \label{fig.outercomp}
        \end{figure}

        Second, we generated the periodogram of the RV before and after we removed the known periodic signals of the three planets using the Lomb-Scargle method, as shown in Figure \ref{fig.outercomp}. In addition to the detected planets, two other peaks at 11 days and 78 days were found at a false-alarm probability (FAP) higher than 0.1\% in the Fourier power. The peak at 11 days was an alias caused by the concentration of the sampling around lunar cycles, as explained in \citet{Christiansen2017}. In the lower panel of Figure \ref{fig.outercomp},  no peak at 11 days was detected, but the peak at 78 days was persistent in both periodograms. The peak around 20 days in the lower panel may be caused by stellar rotation, and the peak around one day was an alias due to data sampling. When we assume a circular orbit with a period of 78 days, this corresponds to a mass of at least 0.03 M$_{Jup}$ for the outer companion, which gives $\beta_{\mathrm{KL}} \simeq 0.5$.  This potential outer companion might explain the high obliquity of HD3167 c if its initial inclination $I_0$ was high enough. 
        
        We found possible indications of an additional outer companion in the system. Additional RV observations of HD3167 on a long time span are necessary to conclusively establish its presence and determine its orbital characteristics, and thus confirm (or refute) our hypothesis.
        
        \section{Conclusion} \label{result}
        We used new observations obtained with HARPS-N to measure the obliquity of a sub-Neptune in a multi-planetary system. The three different methods we applied on this challenging dataset agree, which means that the sky-projected obliquity we measured is reliable. We report a nearly polar orbit for the HD3167 c with $\lambda$ $\sim$ -97$^{\circ}$ $\pm$ 23$^{\circ}$. The measurements of $\lambda$ from RM anomaly, Doppler tomography, and reloaded RM technique agree at better than 1.4~$\sigma$ standard deviation with this value. The $vsini_\star$ from the three methods also agree within their uncertainties. To our knowledge, we are the first to apply these three methods and compared them to the spectroscopic observation of a planetary transit.
        
        These observations are a valuable addition to the known planetary obliquity sample, extending it further beyond hot Jupiters. Several small-radius multi-planet systems with aligned spin-orbits such as Kepler 30 \citep{Ojeda2012} and with a misaligned spin-orbit such as Kepler 56 \citep{Huber2013} have been reported. Additionally, single small exoplanets with high-obliquity measurement such as Kepler 408 \citep{Kamika2019} and GJ436 \citep{Bourrier2018} have also been reported. Some of the misalignments might be explained by the presence of an outer companion in the system. One particularly interesting planetary system is Kepler 56, in which two of its transiting planets are misaligned with respect to the rotation axis of their host star. This misalignment was explained by the presence of a massive non-transiting companion in the system \citep{Huber2013}. A third planet in the Kepler 56 system was later discovered by \citet{Otor2016}. This supported the finding of \citet{Huber2013}. Similarly, the misalignment in HAT-P-11 b may be explained by the presence of HAT-P-11 c \citep{Yee2018}.

        Our dynamical analysis of the system HD3167 places constraints on the obliquity of planet d. We cannot determine the obliquity of planet b with the  current data and information about the system. The Hill-stability criterion shows that the orbits of planets c and d are nearly coplanar, so that both planets are in nearly polar orbits. The interactions of the planets with the stellar spin cannot satisfactorily explain the polar orbits of planets c and d. We postulate that an additional unseen companion exists in the system. This might explain the polar orbits of planets c and d. Indications for additional outer companions are present in the available RV dataset. Continued RV measurements of HD3167 on a longer time span might reveal the outer companion and confirm our speculation.

        \begin{acknowledgements}
                A. P. thanks G. Boué for the helpful discussions during the study. A.C. acknowledges support from the CFisUC strategic project (UID/FIS/04564/2019), ENGAGE SKA (POCI-01-0145-FEDER-022217), and PHOBOS (POCI-01-0145-FEDER-029932), funded by COMPETE 2020 and FCT, Portugal. V. B. acknowledges support by the Swiss National Science Foundation (SNSF) in the frame of the National Centre for Competence in Research PlanetS, and has received funding from the European Research Council (ERC) under the European Union’s Horizon 2020 research and innovation programme (project Four Aces; grant agreement No 724427). S.D acknowledges Vaibhav Pant for insightful discussions.
        \end{acknowledgements}
        
        \bibliographystyle{aa}
        \bibliography{HD3167}
        
        \appendix
        
        \section{Details on the three-vector model}
        \label{app.3vmod}
        
        \subsection{Generic three-vector problem}
        
        The three-vector problem \citep{Boue2006,Boue2014} studies the evolution of the direction of three angular momenta that are represented by the unit vectors $\u_k$ for $k=1,2,3$ in equations
        \begin{align}
        \deriv{\u_1}{t} &= -\nu^{2/1}(\u_1\cdot\u_2)\ \u_2 \times \u_1 - \nu^{3/1}(\u_1\cdot\u_3)\ \u_3 \times \u_1\nnb
        \deriv{\u_2}{t} &=-\nu^{1/2}(\u_1\cdot\u_2)\ \u_1 \times \u_2 -\nu^{3/2}(\u_2\cdot\u_3)\ \u_3 \times \u_2\label{eq.3vpgen}\\
        \deriv{\u_3}{t} &= -\nu^{1/3}(\u_1\cdot\u_3)\ \u_1 \times \u_3-\nu^{2/3}(\u_2\cdot\u_3)\ \u_2 \times \u_3.\nonumber
        \end{align}
        The constants $\nu^{k/j}$ are called the characteristic frequencies and represent the relative influence of the body $k$ over the evolution of $j$. Their expression depends on the considered problem.
        The three-vector problem is integrable \citep{Boue2006}, and the solution is quasi-periodic with two different frequencies.
        Given an initial state where two vectors are aligned and a third is misaligned, it is possible to compute the maximum inclination between the two initially aligned vectors as a function of the initial inclination with the third \citep{Boue2014}.
        The maximum inclination depends on the characteristic frequencies, and the different cases have been classified in section  5.3 of \citet{Boue2014}.

        \subsection{Influence of planet b} 
        \label{app.planetb}
        In Sections \ref{planetmutual} and \ref{sec.tilt}, we claimed that the inclination dynamics of planet b are most likely governed by the star and only influence planets d and c through a modification of the planet-star coupling.
        We present here the justification for this assumption as well as details on the expressions of the coupling constants.
        
        We first focus on the three-vector problem $(\u_\mathrm{S},\u_\mathrm{b} ,\text{and }\u_\mathrm{d})$. For now, we neglect the effect of planet c because we focus on the dynamics of planet b. Following \citet{Boue2006,Boue2014}, the characteristic frequencies that appear in Eq. \ref{eq.3vpgen} are expressed as
        \begin{align}
        \nu^\mathrm{b/S} &= \frac{\alpha_{\mathrm{Sb}}}{S}, & \nu^\mathrm{d/S} &= \frac{\alpha_{\mathrm{Sd}}}{S}  &\nu^\mathrm{b/d} &= \frac{\beta_{\mathrm{bd}}}{G_\mathrm{d}}\nnb
        \nu^\mathrm{S/b} &= \frac{\alpha_{\mathrm{Sb}}}{G_\mathrm{b}},  & \nu^\mathrm{S/d} &= \frac{\alpha_{\mathrm{Sd}}}{G_\mathrm{d}}  & \nu^\mathrm{d/b} &= \frac{\beta_{\mathrm{bd}}}{G_\mathrm{b}},
        \label{eq.freqb}
        \end{align}
        where $G_k$ is the angular momentum of planet $k$,  $S=C_\mathrm{S}\omega_\mathrm{S}$ is the angular momentum of the stellar rotation, with $C_\mathrm{S}$ the stellar moment of inertia and
        \begin{align}
        \alpha_{\mathrm{S}j}& = \frac{3}{2}\frac{\Gr M_\mathrm{S}m_j J_2R_{\star}^2}{a_j^3},\nnb
        \beta_{jk} &=\frac{1}{4}\frac{\Gr m_j m_k a_j}{a_k^2}\mathrm{b}_{3/2}^{(1)}\left(\frac{a_j}{a_k}\right). 
        \label{eq.couplingbeta}
        \end{align}
        Here $\alpha_{\mathrm{S}j}$ represents the coupling between the star and planet $j,$ and $\beta_{jk}$ is the Laplace-Lagrange coupling between planets $j$ and $k$ (we assume $a_j<a_k$). We also define
        \begin{equation}
        J_2 =\frac{k_2\omega_\mathrm{S}^2R_{\star}^3}{3\Gr M_\mathrm{S}},
        \end{equation}
        the gravitational quadrupole coefficient \citep{Lambeck1988}, where $k_2$ is the second fluid Love number of the star and $\omega_\mathrm{S}$ is the stellar rotation speed.
        For the numerical values of $k_2$ and $C_\mathrm{S}$, we use \citep{Landin2009}. For a star of mass $0.85 \mathrm{M}_\odot$, we have ${k_2=0.018}$ and $C_\mathrm{S}/(M_\mathrm{S}R_{\star}^2) =0.10$.
        
        Independently of the stellar rotation speed, we have $\alpha_{\mathrm{Sd}}/\alpha_{\mathrm{Sb}} \leq 0.04$. We therefore neglect the terms depending on $\alpha_{\mathrm{Sd}}$ in this analysis.
        As a result, we can directly apply the results of the analysis reported by \citet{Boue2014}, with the four  characteristic frequencies  $\nu^\mathrm{b/S},\ \nu^\mathrm{S/b},\ \nu^\mathrm{b/d}$ , and  $\nu^\mathrm{d/b}$.
        \begin{figure}
                \includegraphics[width=9cm]{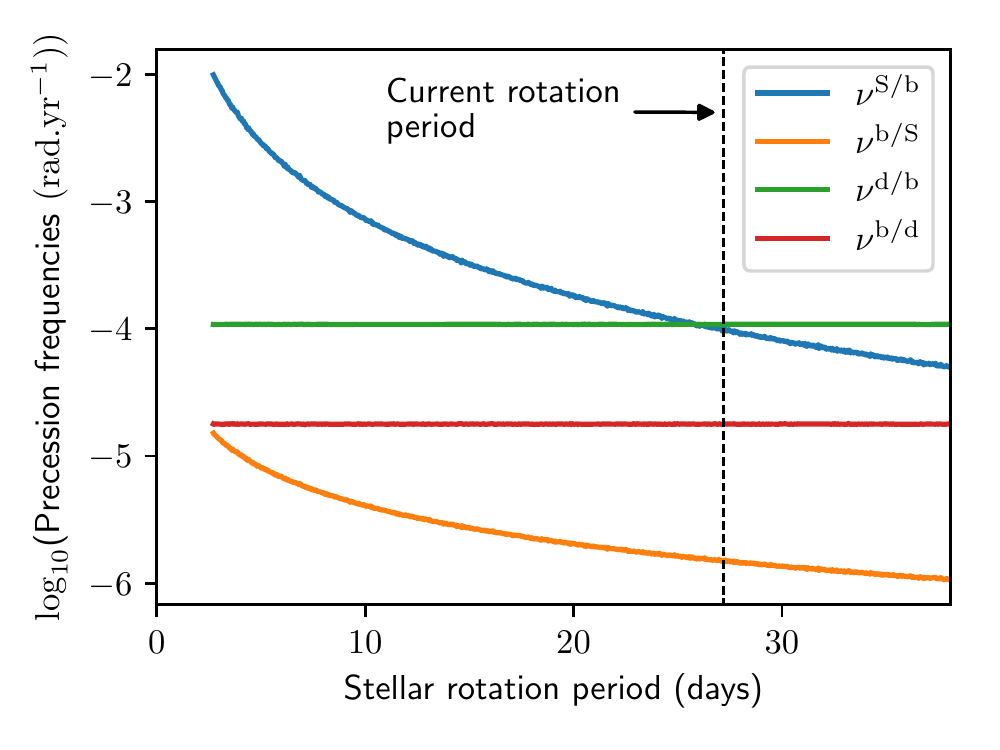}
                \caption{Characteristic frequencies defined in Eq. \eqref{eq.freqb} as a function of the stellar period. The current stellar rotation period is marked with a vertical dashed line. $\nu^\mathrm{b/S}$ dominates for most of the considered frequencies.}
                \label{fig.precb}
        \end{figure}
        
        We plot the frequencies $\nu^{k/j}$ as a function of the stellar period in Figure \ref{fig.precb}. We average the frequencies in each point by randomly drawing the orbital elements from the best fit.

        For the considered range of the stellar revolution period, $\nu^\mathrm{S/b}$ dominates all other frequencies, and it becomes comparable to $\nu^\mathrm{d/b}$ for the current rotation rate. Using the regime classification of \citet{Boue2014}, we can determine the maximum misalignment between $\u_\mathrm{S}$ and $\u_\mathrm{b}$ as a function of the initial inclination between  $\u_\mathrm{S}$ and $\u_\mathrm{d}$.
        
        For a faster-rotating star (i.e., a younger star), we have  ${\nu^\mathrm{S/b}\gg (\nu^\mathrm{b/S},\nu^\mathrm{d/b},\nu^\mathrm{b/d})}$, in which regime no significant misalignment of planet b can be achieved. As a result, planet b is completely coupled with the star and remains within its equator even if the other planets are mutually inclined.
        The current rotation rate leads to the so-called Laplace regime where $(\nu^\mathrm{S/b}\sim\nu^\mathrm{d/b})\gg (\nu^\mathrm{b/S},\nu^\mathrm{b/d}),$ in which the plane of planet b oscillates between the stellar equatorial plane and the plane of planet d. However, if the initial mutual inclination between planet b and the stellar equator is low, planet b remains close to the stellar equator.
        
        We simplify the problem by considering that planet b is coupled to the star and modifies the stellar precession coupling constant $\alpha_{\mathrm{S}k}$ for planets d and c.
        The modification of the coupling constant can be found in \citet[Eq. 129]{Boue2006}\footnote{This equation gives the value of $\tilde\alpha_{\mathrm{S}k}/S,$ but it is straightforward to compute $\tilde\alpha_{\mathrm{S}k}$ from it. }. While the expression was derived for a planet in the presence of a satellite, it remains valid in our case.
        The expression is a generalization of the approximations for close satellites \citep{Tremaine1991} and far satellites \citep{Alembert1749}.
        We denote with $\tilde\alpha_{\mathrm{S}k}$ the modified coupling constant to include the effect of planet b when it is considered as a bulge on the star.
        
        \subsection{Coupling constants for the interactions of the planet with the star}
        \label{app.tilt}
        In section \ref{sec.tilt} we considered the three vectors $\u_{\mathrm{S}},\ \u_{\mathrm{d}}$ , and $ \u_{\mathrm{c}}$.
        The coupling frequencies that appear in Eq. \eqref{eq.3vpgen} for this particular problem are given by 
        \begin{align}
        \nu^\mathrm{d/S} &= \frac{\tilde \alpha_{\mathrm{Sd}}}{S}, & \nu^\mathrm{c/S} &= \frac{\tilde\alpha_{\mathrm{Sc}}}{S}  &\nu^\mathrm{d/c} &= \frac{\beta_{\mathrm{dc}}}{G_\mathrm{c}}\nnb
        \nu^\mathrm{S/d} &= \frac{\tilde\alpha_{\mathrm{Sb}}}{G_\mathrm{d}},  & \nu^\mathrm{S/c} &= \frac{\tilde\alpha_{\mathrm{Sc}}}{G_\mathrm{c}}  & \nu^\mathrm{c/d} &= \frac{\beta_{\mathrm{dc}}}{G_\mathrm{d}},
        \label{eq.freqd}
        \end{align}
        where $\beta_{\mathrm{bd}}$ is defined in Eq. \eqref{eq.couplingbeta} and $\tilde{\alpha}_{\mathrm{S}k}$ is the coupling between the star and planet $k,$ modified to take the influence of planet b into account, as explained in Sect. \ref{app.planetb}.
        We also simplified Eqs. \eqref{eq.3vpgen} by neglecting $\tilde\alpha_{\mathrm{Sc}}$ over $\tilde\alpha_{\mathrm{Sd}}$ because $\tilde\alpha_{\mathrm{Sc}}/\tilde\alpha_{\mathrm{Sd}} <0.05$ independently of the stellar rotation period.
        
        \subsection{Coupling constants for the problem with a companion}
        \label{app.comp}
        
        The characteristic frequencies that govern the evolution of the inclination of the planets under the influence of an outer companion as explained in Sect. \ref{dynamicsouter} are given by 
        \begin{align}
        \nu^\mathrm{pla/S} &= \frac{\alpha}{S} & \nu^\mathrm{pla/comp} &= \frac{\Gamma}{G'}\nnb
        \nu^\mathrm{S/pla} &= \frac{\alpha}{G} & \nu^\mathrm{comp/pla} &= \frac{\Gamma}{G}.
        \label{eq.freqcomp}
        \end{align}
        where
        \begin{align}
        \alpha& =  \sum_{j=\mathrm{b,c,d}}\frac{3}{2}\frac{\Gr M_\mathrm{S}m_j J_2R_{\star}^2}{a_j^3}\nnb
        \Gamma& = \sum_{j=\mathrm{b,c,d}}\frac{3}{4}\frac{\Gr m'm_j a_j^2}{b'^3}.
        \label{eq.couplingcomp}
        \end{align}
        We neglect the interaction between the star and the companion and as a result disregard the corresponding characteristic frequencies.

        \section{Radial velocity data}
        \label{data}
        \begin{table}[H]
                \centering
                \caption{Radial velocities measured on 2017 November 23 with HARPS-N}
                \begin{tabular}{l c c}
                        \hline\hline
                        BJD & RV (ms$^{-1}$)& Uncertainty (ms$^{-1}$) \\ 
                        \hline 

                        58081.30053        &    19534.14    &    1.79    \\
                        58081.31183        &    19534.61    &    1.22   \\
                        58081.3213        &    19534.61    &    2.02    \\
                        58081.34556        &    19532.9        &    0.88    \\
                        58081.35606        &    19536.62    &    0.84   \\
                        58081.36645        &    19534.57    &    0.91   \\
                        58081.3777        &    19538.11    &    0.99   \\
                        58081.38814        &    19536.84    &    0.83   \\
                        58081.3992        &    19536.02    &    0.79   \\
                        58081.40971        &    19534.7        &    0.81   \\
                        58081.42024        &    19535.33    &    0.97   \\
                        58081.43089        &    19533.49    &    1.11   \\
                        58081.44223        &    19532.18    &    0.95   \\
                        58081.45276        &    19533.73    &    0.87   \\
                        58081.46325        &    19533.37    &    0.89   \\
                        58081.47383        &    19532.98    &    0.83   \\
                        58081.48465        &    19532.62    &    0.71   \\
                        58081.49539        &    19533.53    &    0.66   \\
                        58081.5059        &    19531.55    &    0.77   \\
                        58081.51666        &    19530.46    &    1.16    \\
                        58081.52598        &    19530.48    &    1.45   \\
                        58081.53888        &    19532.69    &    2.33   \\
                        58081.54778        &    19530.42    &    1.85   \\
                        58081.55897        &    19527.89    &    2.72    \\

                        \hline 
                        \label{table.2017}
                \end{tabular}
        \end{table}

\end{document}